\documentclass[aps,prl,preprint,groupedaddress,titlepage]{revtex4-1}
\usepackage{graphicx}
\usepackage{gensymb}
\usepackage{amsmath}
\usepackage[colorlinks,hyperindex]{hyperref}
\hypersetup
{
colorlinks,%
citecolor=blue,%
linkcolor=blue,%
urlcolor=blue,%
}

\begin{document}

\title{Multifunctional Nonlocal Metasurfaces}

\author{Adam C. \surname{Overvig}}
\author{Stephanie C. \surname{Malek}}
\author{Nanfang \surname{Yu}}
\email[Correspondence to: ]{ny2214@columbia.edu}
\affiliation{Department of Applied Physics and Applied Mathematics, Columbia University, New York, NY 10027}

\date{\today}

\begin{abstract}
Diffractive photonic devices manipulate light via local and nonlocal optical modes. Local devices, such as metasurfaces, can shape a wavefront at multiple selected wavelengths, but inevitably modify light across the spectrum; nonlocal devices, such as grating filters, offer great frequency selectivity but limited spatial control. Here, we introduce a rational design paradigm using quasi-bound states in the continuum to realize multifunctional nonlocal devices: metasurfaces that produce narrowband spatially tailored wavefronts at multiple selected wavelengths and yet are otherwise transparent.
\end{abstract}

\maketitle

A metasurface is a structured material with one dimension (the out-of-plane direction) comparable to or smaller than the operating wavelength~\cite{yu_light_2011}. The in-plane geometry is a two-dimensional pattern composed of building blocks (called ``meta-units'') chosen so that the collective performs a desired optical functionality. Two broad categories of metasurfaces are (1) local devices that manipulate light via independent scatterers, and (2) nonlocal devices whose response is due to many neighbor-neighbor interactions. Local metasurfaces typically shape optical wavefronts over a broad bandwidth by arranging meta-units on the surface and ignoring nearest neighbor interactions. Nonlocal metasurfaces, in contrast, typically manipulate optical spectra by harnessing the modes supported by many identical adjacent meta-units. 

A prototypical local metasurface is a phase-gradient metasurface~\cite{yu_light_2011,lalanne_blazed_1998,bomzon_space-variant_2002,lalanne_metalenses_2017} deflecting incident light to a desired diffraction order. They are constructed by reference to a library of structures with pre-computed optical responses. Recent advances have enabled multifunctional local metasurfaces, extending single-wavelength, phase-only control to broadband achromatic focusing~\cite{arbabi_controlling_2017,khorasaninejad_achromatic_2017,shrestha_broadband_2018}, complete polarization control\cite{arbabi_dielectric_2015}, and phase-amplitude holography~\cite{wang_broadband_2016,overvig_dielectric_2019}. 

A prototypical nonlocal metasurface~\cite{kwon_nonlocal_2018} is a photonic crystal slab (PCS) supporting a sharp spectral feature (a Fano resonance)~\cite{wang_guided-mode_1990,rosenblatt_resonant_1997,chang-hasnain_high-contrast_2012,khanikaev_fano-resonant_2013,limonov_fano_2017}. Recent study of nonlocal metasurfaces has focused on Bound States in the Continuum (BICs), which are states with infinite radiative Q-factor despite having momentum matched to free-space~\cite{sakoda_symmetry_1995,hsu_observation_2013,bulgakov_bloch_2014,wang_optical_2016,hsu_bound_2016}. In particular, a widely employed approach~\cite{plotnik_experimental_2011,cui_dynamic_2012,kilic_controlling_2008,cui_normal_2016,tittl_imaging-based_2018} applies a perturbation to a PCS supporting a BIC, resulting in an altered, quasi-BIC with a finite Q-factor controllable by the strength of that perturbation, $\delta$\cite{overvig_dimerized_2018,koshelev_asymmetric_2018, fan_temporal_2003}: 
\begin{equation}
Q\propto 1/\delta^2 \label{Qdelt}
\end{equation}
A key advantage of quasi-BICs is simultaneous control of the band structure and Q-factor, enabling devices with compact footprints to retain sharp spectral features: confining light in both space and time~\cite{lemarchand_study_1999,overvig_dimerized_2018,nguyen_symmetry_2018}. 

Nonlocal metasurfaces that spatially shape optical wavefronts have been seldom explored or demonstrated. A nonlocal metasurface with wavefront-shaping properties is a subclass of resonant waveguide grating~\cite{quaranta_recent_2018} designed to diffract light to a desired diffraction order only at frequencies of spatially extended (nonlocal) resonant modes (supermodes) supported by many adjacent meta-units. Such metasurfaces constructed by reference to a rationally designed library of meta-units have not been studied explicitly, though metasurfaces using inverse design achieve somewhat similar behavior~\cite{yang_freeform_2018}. We note that while plasmonic~\cite{yu_light_2011,sun_high-efficiency_2012,ni_metasurface_2013} and Huygens'~\cite{decker_high-efficiency_2015,zhao_dielectric_2016,chen_huygens_2018} metasurfaces are constructed from resonant meta-units, these resonances typically pertain to local modes.

In this Letter, and the accompanying paper~\cite{overvig_selection_2019}, we introduce and utilize an approach employing Group Theory to exhaustively catalog the selection rules governing quasi-BICs. The result is an ``alphabet'' of structures for use as the building blocks of nonlocal metasurfaces. We demonstrate that by using several successive perturbations (more than one letter of the alphabet), multifunctional control of the output spectrum and wavefront is achieved. For instance, we may tailor the Q-factors, polarization angles, and resonant frequencies of up to four modes simultaneously, in principle. By spatially varying the resonant polarization angle (and associated geometric phase) across the device, we demonstrate a wavefront-shaping nonlocal metasurface that spatially shapes circularly polarized light only on resonance and is otherwise completely transparent. We readily extend this capability to a multifunctional nonlocal metasurface that shapes the wavefronts at three wavelengths independently.

We begin by considering a two-dimensional PCS composed of air holes in a silicon slab, sitting on a silicon dioxide substrate. The PCS has square or hexagonal lattice symmetry, supporting quasi-TE and quasi-TM modes whose in-plane symmetries are classified by their irreducible representations~\cite{sakoda_optical_2005}. Group Theory is applied to determine which modes may couple to free-space at normal incidence, and which modes are BICs due to symmetry. As we show in the accompanying paper\cite{overvig_selection_2019}, the condition for coupling free-space light polarized in the $i$-direction to a mode $\psi^0$ via a perturbation $V$, is
\begin{equation}
\Gamma_{\partial_i} = \Gamma_V \otimes \Gamma_{\psi^0}, \label{cc}
\end{equation}
where $\Gamma$ refers to the irreducible representations of the subscripted elements, $\otimes$ denotes the direct product, and $\partial_i$ is the partial spatial derivative in the $i$-direction. For each mode, whose relevant symmetries are specified by $\Gamma_{\psi^0}$, evaluation of Eq.~\ref{cc} tells which polarization, $i$, is coupled (if any) due to a perturbation that breaks in-plane symmetries specified by $\Gamma_V$. By fully classifying the modes existing at the high symmetry points in the unperturbed square and hexagonal lattices, and by exhaustively listing the unique perturbations, $V$, that may be applied to the unperturbed lattice, a full catalog of selection rules may be generated~\cite{overvig_selection_2019}. The modes of a perturbed lattice are classified as:
\begin{equation}
\psi_{L,S}^{m,n},
\end{equation}
where $\psi$ is TM or TE, $L$ is the characteristic reciprocal lattice vector (e.g., $\Gamma$, $M$, or $X$ for a square lattice), $S$ is the irreducible representation (e.g., $A_1$), $m$ is the extended zone order, and $n$ is the out-of-plane order\cite{overvig_selection_2019}. 

\begin{figure}
\includegraphics[width=0.7\columnwidth]{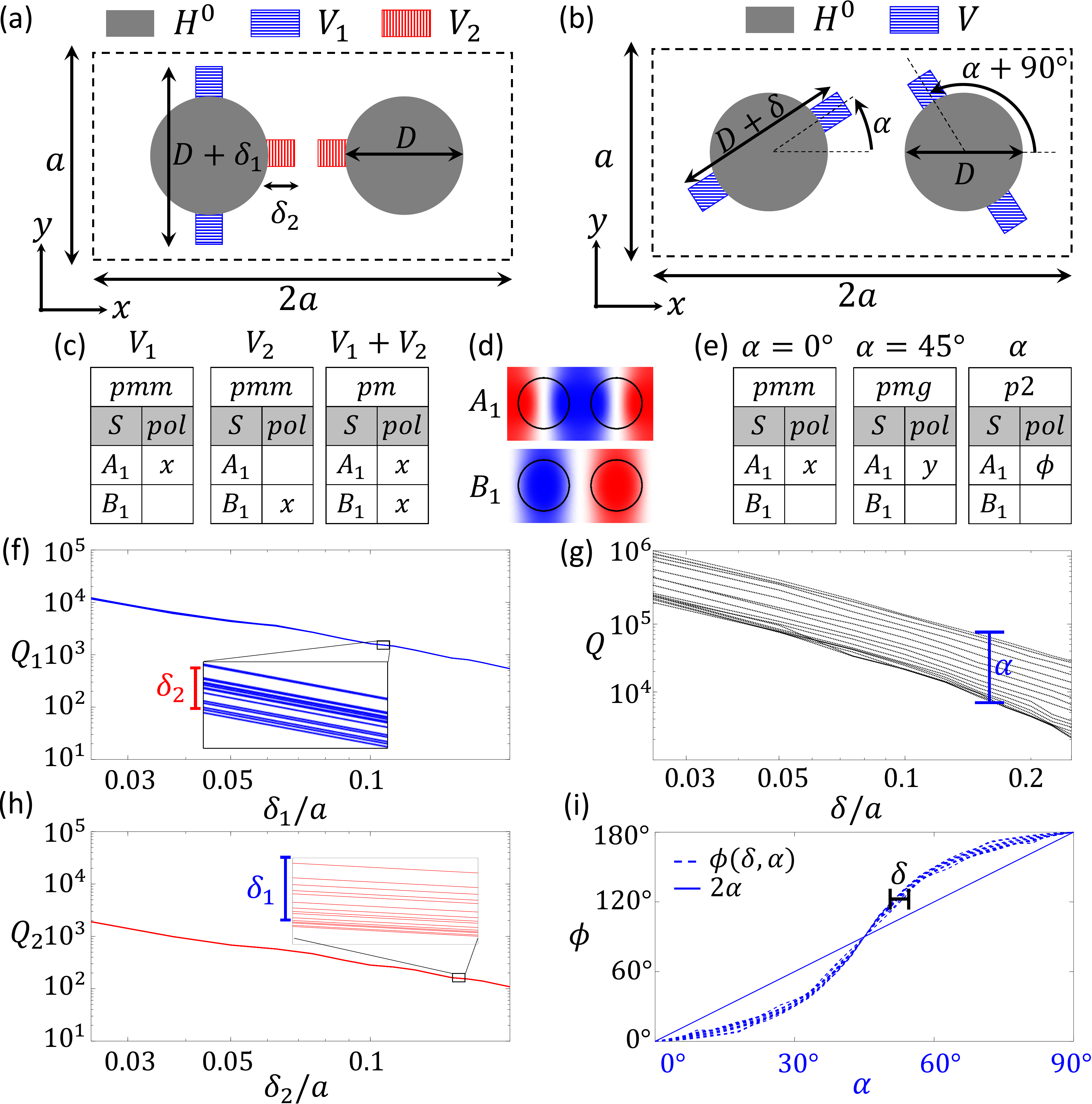}
\caption{\label{fig1} Multifunctional nonlocal metasurfaces. (a) Two perturbations, parameterized by $\delta_1$ and $\delta_2$, are predicted by the selection rules (c) to control the Q-factors of two modes (example field profiles seen in (d)) independently (f,h). (b) A $p2$ space group, characterized by two parameters ($\delta$ and $\alpha$), is predicted by the selection rules (e) to control the Q-factor and polarization angle, $\phi$, of the optical response (g,i).}
\end{figure}

To demonstrate the utility of the resulting alphabet of structures, we identify and explore a few insights evident in the catalog. First, while it has been well-studied recently that a single perturbation may control the radiative lifetime of a single quasi-BIC, we show here that by applying several properly chosen perturbations, several parameters of a light wave may be controlled simultaneously. Figure~\ref{fig1} shows two example cases of multifunctional control. Figure~\ref{fig1}(a) depicts the top-view of a PCS made of Silicon pillars (unperturbed lattice denoted by $H^0$) with two perturbations, $V_1$ and $V_2$, that break distinct symmetries. The selection rules depicted in Fig.~\ref{fig1}(c) predict that these perturbations allow distinct M-point modes (Fig.~\ref{fig1}(d)) to be excited by x-polarized free-space waves: the A1 mode via V1 and the B1 mode via  V2. When both $V_1$ and $V_2$ are applied, the final space group is $pm$, which allows for coupling to both modes~\cite{overvig_selection_2019}. There is therefore a parent-child relationship here: successively applying two ``orthogonal perturbations'' yields a child space group with selection rules inherited from its parents. In this case, characterizing the strength of each perturbation $V_j$ with $\delta_j$ (Fig.~\ref{fig1}(a)), the Q-factors, $Q_j$, of two spectrally separated modes may be controlled independently, as confirmed by fullwave simulations in Figs.~\ref{fig1}(f) and~\ref{fig1}(h).

The second example in Fig.~\ref{fig1} demonstrates control of both Q-factor and polarization angle of the optical response. The space group of the device in Fig.~\ref{fig1}(b) is $p2$, and may be parameterized by two parameters, $\delta$ and $\alpha$. For the choice $\alpha = 0\degree$, the device becomes one parent (with space group $pmm$), while $\alpha = 45\degree$ yields the other parent (with space group $pmg$). The selection rules (Fig.~\ref{fig1}(e)) predict that the $A_1$ mode at the $M$-point is excited by the $x$ and $y$ incident polarizations for the $pmm$ and $pmg$ space groups, respectively, while the intermediate angle $\alpha$ allows coupling to an angle $\phi$ between $x$ and $y$. Figures~\ref{fig1}(g) and~\ref{fig1}(i) confirm this picture, with the relation $\phi \approx 2\alpha$ and $\delta$ controlling the Q-factor. 

The two examples in Fig.~\ref{fig1} demonstrated how the control of quasi-BICs examined through the lens of Group Theory enables multifunctional nonlocal metasurfaces. The accompanying paper~\cite{overvig_selection_2019} explores several applications of the catalog to multifunctional nonlocal metasurfaces, including THz generation using an hexagonal lattice, and mechanically tunable optical lifetimes using a stretchable substrate. Using the two degrees of freedom in Fig.~\ref{fig1}(b) for controlling a single resonance, we now introduce a nonlocal metasurface that spatially shapes an optical wavefront only on resonance. Incorporating the lesson from Fig.~\ref{fig1}(a) that orthogonal pertubations enable control of several resonant modes simultaneously, we then extend this control to three wavelengths.

We begin by studying a geometric phase controlled by the $p2$ space group shown in Fig.~\ref{fig1}(b). It is well-known that a dichroic element imparts a geometric phase $\Phi_{geo} = 2\phi$, where $\phi$ is an eigenpolarization of the element, when an incident right-hand circularly polarized (RCP) light wave is converted into a left-hand circularly polarized (LCP) output. While local metasurfaces (e.g., plasmonic antennas) vary the eigenpolarization according to $\phi=\alpha$, where $\alpha$ is the in-plane orientation angle of the elements, the nonlocal metasurfaces in Fig.~\ref{fig1}(b) vary the eigenpolarization according to $\phi=2\alpha$. (We note that this is not a general rule; the $p2$ perturbation in the $Hex_K$ lattice~\cite{overvig_selection_2019} may be chosen such that $\phi=-4\alpha$, for instance.) Consequently, for RCP incidence, the phases of transmitted LCP and reflected RCP light vary as $\Phi_{geo}\approx4\alpha$, while RCP light in transmission and LCP light in reflection are invariant to $\alpha$. 

\begin{figure*}
\includegraphics[width=1\columnwidth]{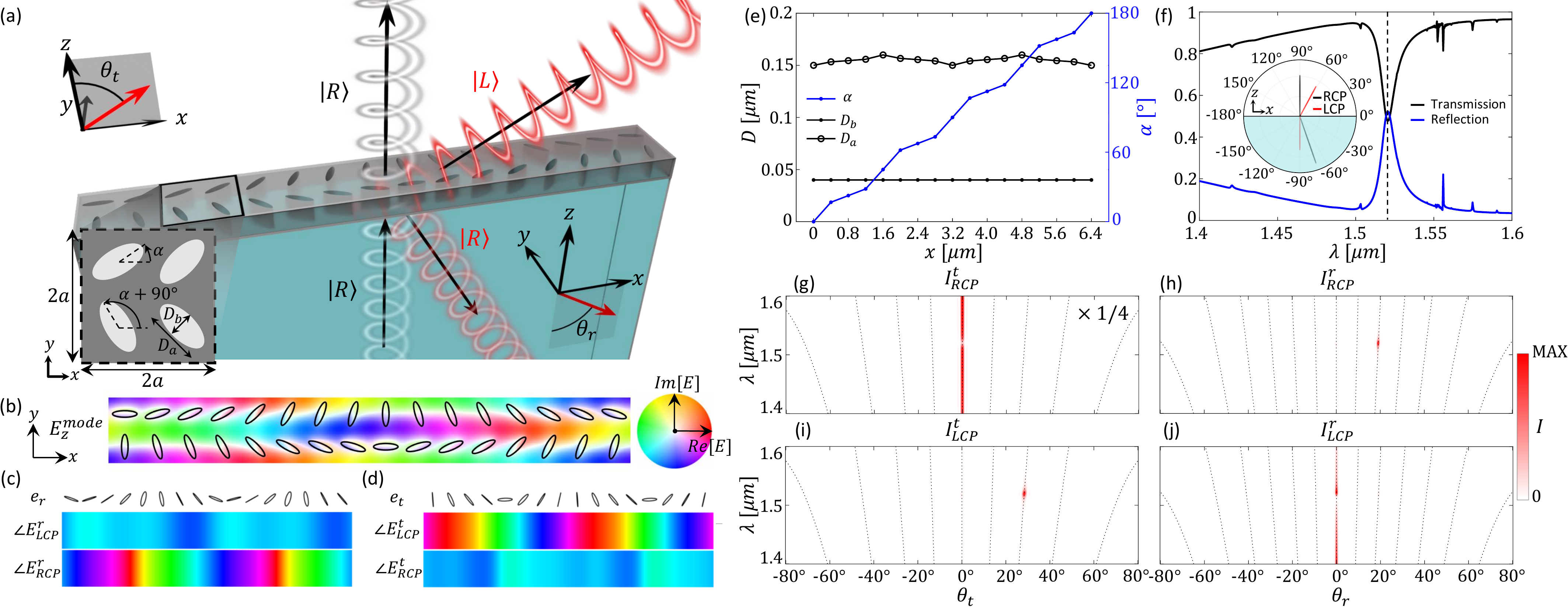}
\caption{\label{fig2} Nonlocal phase gradient metasurface. (a) Schematic depicting the device: a $250nm$ thin film of Silicon on top of quartz is patterned with elliptical holes resonantly deflecting light only in a narrow spectral band (red) when excited by broadband circularly polarized light (white). (b) Top-view of the resonant metasurface overlaid with the complex field on resonance. (c,d) Top-view of spatial profiles for the reflection (transmission) side polarization state, $e_r$ ($e_t$), phase of the RCP component, $E_{RCP}^r$ ($E_{RCP}^t$), and phase of the LCP component, $E_{LCP}^r$ ($E_{LCP}^t$). (e) Geometrical parameters for the device in (b). (f) Transmission and reflection spectra of the device in (b), with inset showing the far-field projection on resonance. (g-j) Far-field projections of each component, showing deflection of light with converted handedness only on resonance.}
\end{figure*}

We may therefore vary the phase of the resonantly converted light, while the non-resonant light is unaffected to first order. Figure~\ref{fig2}(a) schematically depicts such a metasurface consisting of a slab of Silicon etched with elliptical holes that encode a phase gradient. When the metasurface is excited by a broadband RCP wave at normal incidence from the glass substrate, converted light waves at the resonant frequency are anomalously deflected. The field on resonance (Fig.~\ref{fig2}(b)) is that of a supermode of the entire structure, which is slightly blueshifted from the resonant frequency in the case of no phase gradient~\cite{overvig_selection_2019}. Figure~\ref{fig2}(c) (Fig.~\ref{fig2}(d)) shows the spatially tailored reflected (transmitted) polarization state, and the phase profiles of the reflected (transmitted) circularly polarized components. Figure~\ref{fig2}(e) shows the geometric parameters (see the inset in Fig.~\ref{fig2}(a)) along the metasurface, where $\alpha$ at each position is chosen to linearly grade the corresponding geometric phase (Fig.\ref{fig1}(i)), and the semi-major diameter of the ellipse is varied to maintain a constant resonant frequency. 

The spectra of the device, depicted in Fig.~\ref{fig2}(f), show a primary peak associated with the resonant mode. The inset depicts the far-field projection of the exiting LCP and RCP components at the resonant wavelength, confirming that deflection only occurs for light with the converted handedness. Figures~\ref{fig2}(g-j) depict the far-field at each wavelength near the resonance, showing that there is only deflection on resonance. Notably, because $\alpha$ has varied through $180\degree$, the phase has varied across $4\pi$, and so the diffraction order is $m = 2$ (see the overlaid diffraction orders in Fig.~\ref{fig2}(h,i) (dashed contours)). 

\begin{figure}
\includegraphics[width=0.7\columnwidth]{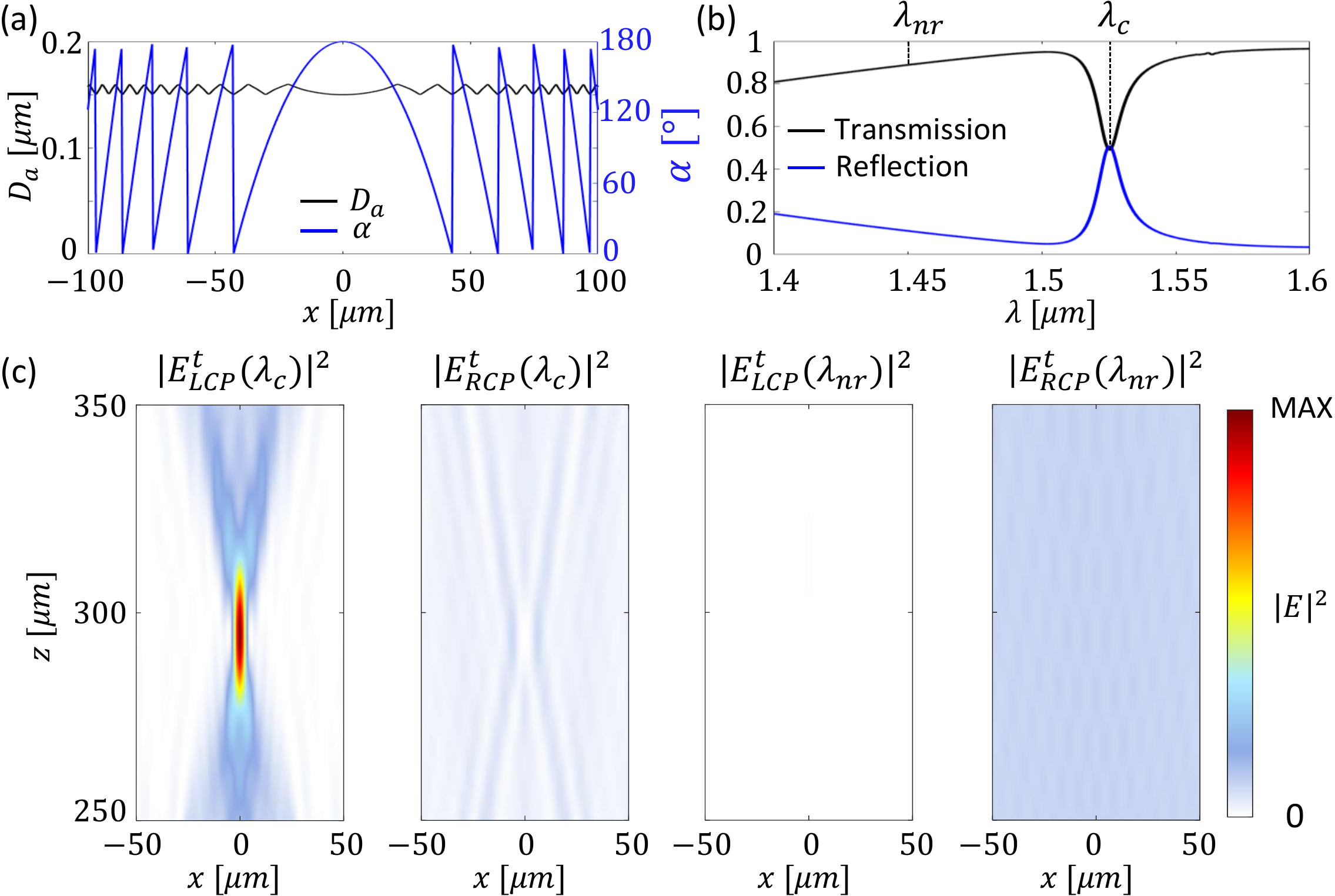}
\caption{\label{fig3} Nonlocal metasurface lens. (a) Geometrical parameters varying across the device. (b) Transmission and reflection spectra of the entire device. (c) Far-field distributions near the designed focal spot on the transmission side, showing the LCP and RCP intensities at the central wavelength of the resonance, $\lambda_c$, and at a non-resonant wavelength, $\lambda_{nr}$. }
\end{figure}

Next, Fig.~\ref{fig3} demonstrates more complex spectrally selective wavefront shaping. Figure~\ref{fig3}(a) shows the geometric parameters of the elliptical holes for a nonlocal metasurface lens. The resonance remains intact (Fig.~\ref{fig3}(b)) despite the variation of the geometry along the device. The resulting LCP and RCP far-field distributions at two wavelengths are shown in Fig.~\ref{fig3}(c), confirming that focusing occurs only on resonance for the circularly polarized component with converted handedness. We note that the devices in Fig.~\ref{fig2} and Fig.~\ref{fig3} may shape the wavefront in the $y$-direction as well~\cite{overvig_selection_2019}, and that the device in Fig.~\ref{fig3} may prove useful for augmented reality applications as compact and highly transparent lenses.

The devices in Figs.~\ref{fig2} and~\ref{fig3} represent a major departure from Fano resonances traditionally studied in nonlocal metasurfaces. In order to ensure complete interference of the bright and dark modes, nonlocal metasurfaces are conventionally two-port systems (i.e., only the $m=0$ diffraction order in the transmission and reflection sides exist)~\cite{fan_temporal_2003}. In contrast, these nonlocal metasurfaces demonstrate that Fano resonances may occur due to interference involving any diffraction angle (the $m=\pm2$ diffraction order), with power exiting to only four ports of the system. All of the remaining diffraction orders present in the devices (dashed contours in Fig.\ref{fig2}(g-j)) are excluded as interference pathways due to in-plane momentum conservation in the presence of the phase gradient, reminiscent of the Generalized Snell's Law in local metasurfaces~\cite{yu_light_2011}. In nonlocal metasurfaces, however, momentum conservation is enforced both upon coupling in and upon coupling out: light resonantly couples to a supermode (see Fig.~\ref{fig2}(b)) with a Bloch wavevector $k_B=2\pi/P$, where $P=16a$ is the superperiod, and then couples out to the reflection port with an additional factor $\pm k_B$, where the sign is determined by the handedness of the light. That is, light can only reflect to the $m=0$ and $m=2$ diffraction orders. Chirality (and therefore spin-selectivity) is required to eliminate the LCP pathways on each side for RCP incidence, and is the subject of future work.

Due to the mediation by a supermode, the resonant deflection occurs at the frequency of this traveling-wave supermode rather than that of the standing-wave mode at the $\Gamma$-point. The difference in these frequencies follows the angular dispersion of the supermode (its band structure)\cite{overvig_selection_2019}. A device with varying deflection angle, such as the lens in Fig.\ref{fig3}, must be designed with this angular dispersion in mind: the frequency shift across the device must be smaller than the resonant linewidth. For instance, for a dispersion following $\omega_{res}(k) = \omega_{0} + bk^2$, where $k=k_0 sin(\theta_t)$, $k_0$ is the free-space wavevector, $\omega_0$ is the angular frequency of the mode at $k = 0$, and $b$ is a constant, the nonlocal lens should satisfy
\begin{equation}
 \text{NA}^2 \leq \frac{\omega_0/ k^2_0}{\left \lvert b \right \rvert Q}, \label{Qbk}
\end{equation}
where $\text{NA}$ is the numerical aperture. This constraint suggests that high NA lenses require flat bands ($\left \lvert b \right \rvert$ should be small). Alternatively, the resonant frequency of the supermode should be spatially adjusted through meta-unit design to counteract the shift. A key strength of an approach using quasi-BICs is that the functionality is due to a small perturbation, leaving the unperturbed geometrical parameters free to tune. This enables a design paradigm that optimizes the unperturbed structure considering resonant frequency and band curvature, and then applies the perturbation (guided by the catalog) to realize the wavefront shaping functionality~\cite{overvig_selection_2019}.

\begin{figure}
\includegraphics[width=0.7\columnwidth]{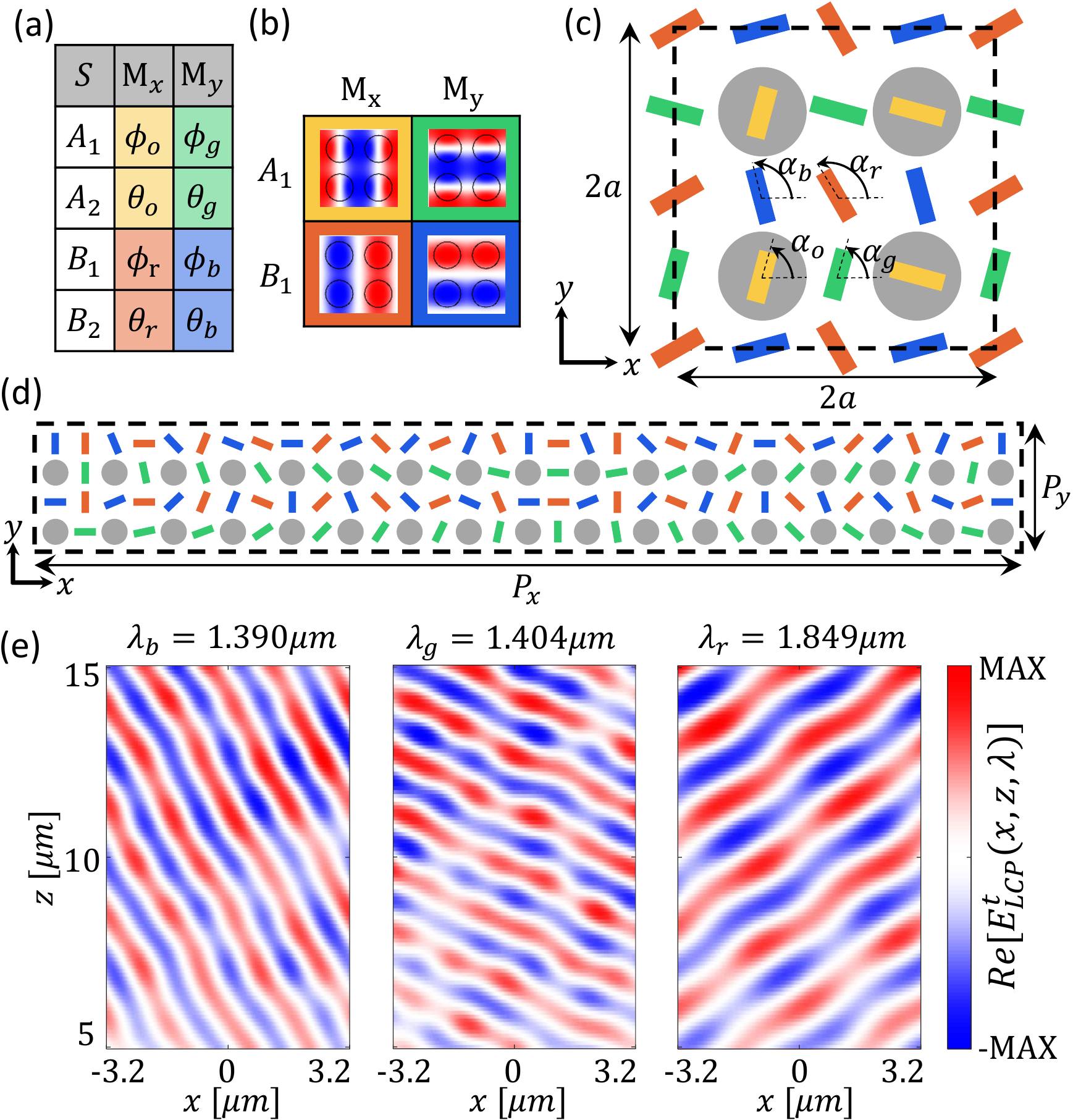}
\caption{\label{fig4}Multi-wavelength resonant metasurfaces. (a) Selection rules of a space group targeting four independent mode symmetries (examples given in (b)), with a meta-unit shown in (c). (d) Top-view of a superperiod of a device (where features denote holes etched in a $250nm$ Silicon slab) imparting distinct phase gradients to three wavelengths simultaneously. $P_x = 6.4\mu m$ and $P_y = 0.8\mu m$, the circles have diameters of $180nm$ and the rectangles all have dimensions $50nm \times 150nm$ . (e) Electric field of the LCP component transmitted at three resonant wavelengths, demonstrating independent control of three supermodes.}
\end{figure}

Finally, we extend the functionality of nonlocal metasurfaces by successively adding perturbations, each capable of controlling the linewidth and geometric phase of separate modes. In particular, we identify four orthogonal $p2$ space groups targeting four distinct mode symmetries, allowing control of the Q-factor and polarization angle of each. Figure~\ref{fig4}(a) shows the selection rules of the four independently controlled modes, with example mode profiles shown in Fig.~\ref{fig4}(b), and a meta-unit shown in Fig.~\ref{fig4}(c). By spatially varying the geometric phase of each mode, a device with the nonlocal functionality demonstrated in Fig.~\ref{fig2} may be realized with four distinct phase profiles encoded by the four distinct perturbations. As a proof-of-principle, we apply three such perturbations, shown in Fig.~\ref{fig4}(d). Fullwave simulations confirm that, at the corresponding resonant wavelengths, three different phase gradients have been faithfully encoded for the output light, deflecting light with converted handedness of circular polarization to independent far-field angles (Fig.~\ref{fig4}(e)). While in principle four distinct profiles should be possible, we find that cross-talk (either second-order or due to the traveling wave supermodes having slightly distinct selection rules to their standing-wave counterparts) degrades the performance compared to excluding the fourth. The multifunctional nonlocal metasurface in Fig.~\ref{fig4} demonstrates that nonlocal metasurfaces are capable of independently controlling at least six parameters of an optical spectrum (Q-factors and polarization angles of three resonances), not counting resonant wavelengths and angular dispersion. Further work is needed to reduce the cross-talk to improve the performance of multifunctional nonlocal metasurfaces. 

In summary, we introduced an approach based on Group Theory to study and catalog the selection rules of quasi-BICs controlled by in-plane perturbations to PCSs. In this Letter, we used the resulting ``alphabet'' of structures to design nonlocal metasurfaces with multifunctional spectral control, and we introduced a nonlocal metasurface that spatially shapes the outgoing wavefront only at the designed resonant frequencies. By combining these two insights, we demonstrated a multifunctional wavefront-shaping nonlocal metasurface, and demonstrated that the catalog of selection rules provides a rich starting point for research into nonlocal optical devices. The exclusive command of narrow resonant bands in otherwise entirely transparent devices are highly promising for augmented reality applications, and the enhanced light-matter interactions due to modal overlap and large, designer Q-factors suggest that multifunctional nonlocal metasurfaces are a platform uniquely suitable for expanding capabilities in active and nonlinear optics. 

\begin{acknowledgments}
The authors thank Andrea Al\`{u} for helpful discussions. The work was supported by the Defense Advanced Research Projects Agency (grant nos. D15AP00111 and HR0011-17-2-0017), the National Science Foundation (grant nos. ECCS-2004685 and QII-TAQS-1936359), and the Air Force Office of Scientific Research (grant no. FA9550-14-1-0389). A.C.O. acknowledges support from the NSF IGERT program (grant no. DGE-1069240). S.C.M acknowledges support from the NSF Graduate Research Fellowship Program (grant no. DGE-1644869).
\end{acknowledgments}

%

\end{document}